\documentclass[10pt,letterpaper,twocolumn]{article} 

\newcommand{\be}{\begin{equation}}
\newcommand{\ee}{\end{equation}}
\newcommand{\bea}{\begin{eqnarray}}
\newcommand{\eea}{\end{eqnarray}}
\newcommand{\beaa}{\begin{eqnarray*}}
\newcommand{\eeaa}{\end{eqnarray*}}
\newcommand{\ben}{\begin{enumerate}}
\newcommand{\een}{\end{enumerate}}
\newcommand{\bi}{\begin{itemize}}
\newcommand{\ei}{\end{itemize}}

\newcommand{\lip}{\langle}
\newcommand{\rip}{\rangle}
\newcommand{\uu}{\underline}

\newcommand{\df}{{\rm d}}
\usepackage{ol2}
\usepackage[draft]{hyperref}
\usepackage{amsmath}

\begin{document}

\twocolumn[ 

\title{Loss of polarization entanglement in a fiber-optic system with polarization mode dispersion in one optical path}

\author{Misha Brodsky$^{1,*}$, Elizabeth C. George$^{1}$, Cristian Antonelli$^2$ and Mark Shtaif$^3$}
\address{$^1$AT\&T Labs, Middletown, NJ 07748 USA \\ $^2$Dipartimento di Ingegneria Elettrica e dell'Informazione and CNISM, Universit\`a dell'Aquila, 67040 L'Aquila, Italy \\ $^3$Department of Electrical Engineering, Physical Electronics, Tel-Aviv University, Tel-Aviv 69978, Israel \\ $^*$Corresponding author: brodsky@research.att.com }

\begin{abstract}
We characterize theoretically and experimentally the degradation of polarization entanglement in a fiber-optic entanglement distribution system where one of the optical fibers is exposed to the effects of polarization mode dispersion (PMD). We show gradual reduction of entanglement with increasing PMD and find that the highest PMD tolerance is achieved when the bandwidth of the pump used to generate the entangled photons in a $\chi^{(3)}$ process is approximately half the bandwidth of the quantum channels.
\end{abstract}


] 

\noindent

Entanglement is a key ingredient in quantum communications and computing, with one of its most commonly considered applications being the exchange of cryptographic keys between remote parties \cite{Ekert}. Almost all such applications strongly rely on the ability to create and distribute entanglement between distant locations. By applying quantum entanglement to pairs of photons, these applications can be naturally integrated into existing wavelength division multiplexed (WDM) networks, thereby taking advantage of the vast fiber-optic infrastructure, which is ubiquitously deployed around the globe. Among several schemes proposed to entangle a pair of photons \cite{Gisin07}, polarization entanglement is particularly appealing because of the ease with which polarizations can be manipulated using standard optical instrumentation. However, it was suggested previously that sufficiently large polarization mode dispersion (PMD) could severely deteriorate the entanglement between two photons \cite{Gisin02}. Interestingly, recent experiments demonstrated successful distribution of polarization entanglement, in which only one photon of the pair traveled over a long fiber \cite{PoppeECOC06, Hubel07, KumarPDP}. The PMD was not measured in those experiments and was presumed to be very low.

In this paper we address the question of experimentally assessing the tolerance of applications involving polarization entanglement of photons to the effects of PMD. In this first set of experiments, we focus on an illustrative case where only one of the two photons is affected by PMD - exactly the same configuration as in \cite{PoppeECOC06, Hubel07}. The entangled photons are distributed between the two legitimate users Alice and Bob. While Alice is located close to the entanglement source, Bob's station is at a notable distance from it. His photon is sent to him over a WDM channel and is thereby exposed to the effects of PMD \cite{PoppeECOC06,Hubel07}. We show that, in contrast to the more general case in which both photons are exposed to the effects of PMD \cite{Brodsky_ECOC2010}, the decay of entanglement with increasing PMD is gradual, i.e. the entanglement sudden death phenomenon \cite{Sd_Yu} does not occur. We quantitatively characterized the role of the pump bandwidth in the tolerance of the entanglement to PMD, and find that the optimal value of the pump is approximately half of the bandwidth of the filters used for selecting the channels of the WDM system. In contrast to classical communications, the entanglement degradation due to PMD in one optical path does not depend on the orientation of the PMD vector. 

\begin{figure}
\begin{centering}
\includegraphics[width=.95\columnwidth]{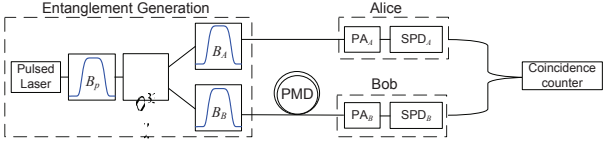}
\caption{Experimental setup schematic.} \label{Setup}
\end{centering}
\end{figure}

A sketch of the experimental setup is shown in Fig. \ref{Setup}. For generation and detection of entangled photon pairs we employ an entangled photon transmission system custom-built by NuCrypt \cite{Nucrypt}. The details of the entangled pair creation are described by Wang et. al. in \cite{WangOFC09, WangQE09}. In brief, by converting light from a tunable pulsed pump we first create a polarization-entangled state described by the quantum state vector
\be | \psi_p \rip = \left( |\uu h,\uu h\rip  + \exp(i\alpha) |\uu v,\uu v\rip \right)/\sqrt 2 \ee
where $\uu h$ and $\uu v$ are Jones vectors in the reference frame of the source and $\alpha$ is an arbitrary fixed phase factor.
The pump pulses are generated by filtering the output of a pulsed laser source with a band-pass filter, whose bandwidth we denote by $B_p$. In essence, since the pulsed laser output consists of ultra-short pulses, $B_p$ can be considered as the actual bandwidth of the pump. In the set of experiments presented here the 3dB bandwidth of the pump was set to $B_p=75$GHz and to $B_p=120$GHz. Then we spectrally separate the two photons with WDM filters of the type that is used for optical add-drop multiplexing ($B_A$ and $B_B$ in Fig.1) positioned at 193.4THz and 192.4THz. The optical filters determine the individual photon spectra and serve also to direct the photons to two optical fibers. In the experiment two filter bandwidths were used for all purposes, $B_A= B_B = 130$GHz or $B_A = B_B = 70$GHz. The power transmittivities of all filters was found to resemble very closely a third-order super-Gaussian function, a property that we exploit in the comparison with the theoretical analysis. In our scheme, one path has a negligible differential group delay (DGD) whereas the DGD value in the second path is being increased incrementally by inserting various combinations of concatenated polarization maintaining (PM) fiber jumpers each with known DGD. The principal states of the individual PM jumpers are aligned to each other so that the overall DGD is the sum of the individual DGDs. The DGD of each configuration was also measured by using an interferometric technique. Each fiber is fed into a polarization analyzer (PA in Fig. 1), which consisted of a polarization controller and a polarizer, and was subsequently connected to single photon detectors (SPD). The SPDs efficiencies are about 20\% and afterpulsing is removed digitally. By rotating the analyzers and recording the coincidental detections we perform complete quantum state tomography \cite{Tomochapter} for each value of DGD, thus experimentally obtaining the polarization density matrix, which completely characterizes the two-photon state. At a repetition rate of 45MHz it takes about 1 hour to complete one measurement (one density matrix).
To quantify the entanglement of the received two-photon state we use concurrence $C$ \cite{Wootters}, which is the most ubiquitously used entanglement measure for two-qubit systems. In addition, we also compute the \textit{maximum possible} $S$ parameter, which when greater than 2 indicates violation of Bells inequality in the Clauser, Horne, Shimony, and Holt definition \cite{CHSH}.

%

The quantum state of the generated photon pair, which thereinafter we refer to as the input state, is expressed as the tensorial product of two terms,$ | \psi_{\mathrm{in}} \rip = | \psi_p \rip \otimes | g(t_A,t_B) \rip$, where $| \psi_p \rip$ describes the polarization properties of the generated state and the term $| g(t_A,t_B) \rip$ describes its time content. Since the entanglement is achieved via the $\chi^{(3)}$ process, $| g(t_A,t_B) \rip$ is expressed as 
\bea |g(t_A,t_B)\rip &=& \int\df t h_A^*(t-t_A)h_B^*(t-t_B) \nonumber \\ &\times& \int \df t_p E_p(t-t_p) E_p(t_p) | t_A,t_B \rip, \label{700} \eea
where $h_A(t)$ and $h_B(t)$ are the respective inverse Fourier transforms of $H_A(\omega)$ and $H_B(\omega)$, which are the transfer functions of two frequency filters positioned along Alice and Bob's optical paths.  The term $E_p(t)$ is the complex time envelope of the pump signal. It has been assumed in the derivation of Eq. (\ref{700}) that the phase matching condition is uniformly satisfied within the bandwidth of each of the two filters, so that the spectral dependence of phase matching is negligible.

Given that PMD is present in only one optical path, the polarization dependent part of the input state can be re-expressed as $ |\psi_p \rip = \left( |\uu s , \uu b \rip + |\uu s' ,  \uu b' \rip \right) / \sqrt 2$, where $\uu s$ and $\uu s'$ are the slow and the fast principal states of polarization (PSP) respectively, and the Jones vectors $\uu b$ and $\uu b'$ are given by $\uu b = (\uu s \cdot \uu h) \uu h  + e^{i\alpha} (\uu s \cdot \uu v) \uu v$ and $\uu b' =  (\uu s' \cdot \uu h) \uu h + e^{i\alpha} (\uu s' \cdot \uu v) \uu v$ (we use primes to denote orthogonality in Jones space). Note that the phase factor $\alpha$ has been removed and the polarization input state in this representation is a standard Bell $\Phi^+$ state. Use of this basis allows us to represent the two-photon state received by Alice and Bob, which we call the \textit{output state}, in the simple form
\bea |\psi_{\mathrm{out}} \rip &=& \frac{1}{\sqrt 2}  \left[ |\uu s , \uu b \rip \otimes |g(t_A-\tau/2,t_B) \rip \right. \nonumber \\  &+& \left. |\uu s' ,  \uu b' \rip \otimes |g(t_A+\tau/2,t_B)  \rip \right].  \label{800}\eea
The density matrix $\rho$ characterizing the polarization properties of the output state is obtained by tracing the full density matrix $|\psi_{\mathrm{out}}\rip\lip\psi_{\mathrm{out}}|$ over the time modes,
that is $ \rho=\int\df t'_A\df t'_B \lip t'_A,t'_B|\psi_{\mathrm{out}}\rip\lip\psi_{\mathrm{out}}|t'_A,t'_B\rip$. Here tracing over the time modes accounts for the
insensitivity of the receivers to the photon's time of arrival. The only nonzero terms in the resultant density matrix are $\rho_{11}=\rho_{44} = 1/2$ and $\rho_{14}=\rho_{41}^* = R(\tau)/2$, where
\bea R(\tau) &=& \kappa \int \int \df \omega_A \df \omega_B |H_A(\omega_A)|^2 |H_B(\omega_B)|^2\nonumber \\ &\times&  \left| \tilde E_p\left( \frac{\omega_A + \omega_B}{2} \right) \right|^4 e^{i\tau\omega_A}, \label{1000} \eea
with $\tilde E_p(\omega)$ being the Fourier transform of $E_p(t)$ and $\kappa$ being a normalization coefficient such that $R(0)=1$. For the density matrix described above one can readily calculate the concurrence $C = |R(\tau)|$. In addition, applying Horodecki's procedure \cite{Horodecky}, the maximum violation of Bell's inequality can be shown to be $S = 2 \sqrt{1 + |R(\tau)|^2} = 2 \sqrt{1 + C^2}$. Note that $C=1$ and $S = 2 \sqrt 2$ when $\tau=0$.

An idea on the correspondence between theory and experiments can be obtained by calculating the fidelity between the measured and the theoretically predicted density matrices for increasing DGD values. The calculated fidelity ranges between 0.94 and 0.99.

\begin{figure}
\begin{centering}
\includegraphics[width=.95\columnwidth]{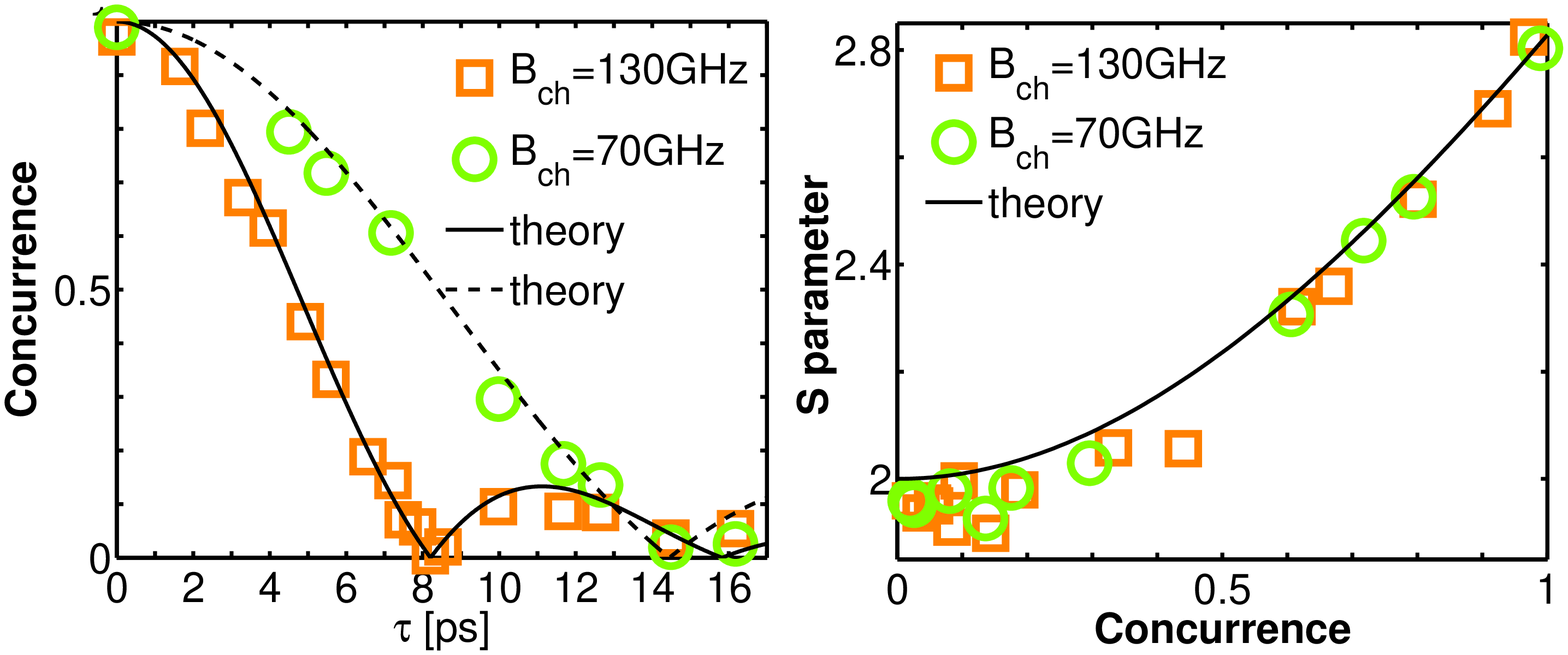}
\caption{(a) Concurrence versus DGD. (b) Bell $S$ parameter versus concurrence $C$. Squares correspond to $B_A=B_B = B_{\mathrm{ch}} =130$GHz and $B_p = 120$GHz, circles to $B_A=B_B= B_{\mathrm{ch}} = 70$GHz and $B_p = 75$GHz.} \label{Exp2}
\end{centering}
\end{figure}

To further quantify the agreement between experiment and theory we compute concurrence from the experimental density matrices and plot it in Fig. \ref{Exp2}a together with the expression $C = |R(\tau)|$, in which the filter transfer functions are third order super-gaussians with a FWHM equal to that of the experimental filters.
The symbols (square and circles) represent results obtained with the two filter bandwidths that were available to us in this experiment.
The theoretical and experimental results nearly coincide. The data has very small statistical errors, as the size of the error bars would be similar to that of the symbols used to plot the figures.
As is evident from Fig. \ref{Exp2}a, the sensitivity to increasing the DGD is smaller for lower channel bandwidth. Indeed narrowing the photons' bandwidths increases their time uncertainty, thus reducing the sensitivity to the effects of PMD. We show in Fig. \ref{Exp2}b the $S$ parameters as a function of the concurrence, illustrating both the experimental results that were extracted from the measured density matrices and the theoretical relation  $S = 2\sqrt{1+C^2}$.  Interestingly, when PMD is present in one fiber only the $S$ parameter remains greater than 2 for any PMD value, thus making nonlocality-based quantum protocols \cite{Acin_PRL07}, at least in principle, viable for arbitrarily large PMD values. Notice that $S$ calculated from experimental density matrices falls somewhat below the theoretical curve for small values of $S$, which corresponds to large DGD values. These large values required the concatenation of many PM jumpers. We speculate that unavoidable misalignment between consecutive PM jumpers produced some high-order PMD effect that brought down the experimental $S$ values.

\begin{figure}
\begin{centering}
\includegraphics[width=.95\columnwidth]{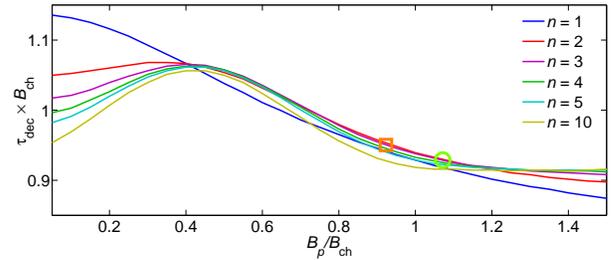}
\caption{Normalized DGD value $\tau_{\mathrm{dec}}$ versus the ratio $B_p/B_{\mathrm{ch}}$, for $B_A = B_B =  B_{\mathrm{ch}}$. The different curves correspond to super-Gaussian filters of various orders $n$. The symbols, to be interpreted as in Fig. \ref{Exp2}, show available experimental points.} \label{tau_th}
\end{centering}
\end{figure}

It then remains to find out the dependence of the PMD tolerance to the bandwidth of the pump signal. We quantify the robustness to PMD in terms of the amount of DGD that is needed for the concurrence to drop by 90\% to the level of $C=0.1$. We denote the DGD value that causes such drop in concurrence by $\tau_{\mathrm{dec}}$, with the subscript $\mathrm{dec}$ standing for ``decoherence." In Fig. \ref{tau_th} we plot $\tau_{\mathrm{dec}}$ normalized to $B_{\mathrm{ch}}^{-1}$ versus the ratio $B_p/B_{\mathrm{ch}}$, for $B_A = B_B =  B_{\mathrm{ch}}$. Each curve is calculated by modeling the filter transmittivities with super-Gaussian functions of order $n$. The plot shows that, in the case of gaussian functions ($n=1$) an increase in the pump bandwidth always leads to a decrease in $\tau_{\mathrm{dec}}$. Yet, for $n>1$ there exists a value of $B_p$ at which $\tau_{\mathrm{dec}}$ is largest. The optimal value of $B_p$ ranges between $0.4B_{\mathrm{ch}}$ and $0.5 B_{\mathrm{ch}}$, depending on the order of the super-Gaussian filter. Apparently, the existence of an optimal pump bandwidth is related to the squarish shape of the WDM filter transmittivities. We have verified that the existence of an optimal pump bandwidth persists also in the case when only the channel filters are squarish super-Gaussian, while the shape of the pump filter is Gaussian.

To conclude, we have characterized experimentally and theoretically the effects of PMD on the distribution of polarization entangled photons between two users of an fiber-communications system. In our configuration, which models recent field demonstrations of distant entanglement \cite{PoppeECOC06,Hubel07},
only one photon experiences PMD effects. Our study shows the gradual disappearance of entanglement with increasing values of DGD, contrary to the more general case where PMD is present in both arms \cite{Brodsky_ECOC2010}. Maximum PMD tolerance corresponds to the case where the bandwidth of the pump used to generate the entangled photon-pair is equal to approximately one half of the bandwidths of the quantum channel filters.

\end{document}